\documentclass[a4paper]{article}

\usepackage{a4wide,amsmath,amsthm}
\usepackage{amssymb}
\usepackage{enumitem}
\usepackage{algorithm,algorithmic}
\usepackage{xcolor}
\usepackage{tikz}
\usetikzlibrary{decorations.pathreplacing}
\usetikzlibrary{arrows.meta}
\usetikzlibrary{intersections}
\usetikzlibrary{bending}
\usetikzlibrary{backgrounds}
\usetikzlibrary{shapes.geometric}
\usetikzlibrary{shapes.misc}
\usetikzlibrary{calc}
\usepackage{hyperref}

\newtheorem{theorem}{Theorem}
\newtheorem{lemma}[theorem]{Lemma}
\newtheorem{invariant}[theorem]{Invariant}

\newlength{\halflineskip}
\newcommand{\rightbrace}[3][0]{\hspace{0pt plus 1filll}\smash{\tikz[baseline=#1\baselineskip]{\draw[thick,color=gray,line cap=round,decorate,decoration={brace,amplitude=2.75pt}] (0pt,0.7\baselineskip) -- ++(0pt,-#2\baselineskip+1pt) (0,\halflineskip-#2\halflineskip) node[color=gray,right,inner sep=5pt,anchor=base west]{\makebox[3.3cm][l]{#3}};}}}

\algsetup{indent=1.5em,linenodelimiter=}
\newcommand{\BEGIN}[1][]{\LOOP[#1]}\newcommand{\END}{\ENDLOOP}
\renewcommand{\algorithmicloop}{\textbf{begin}}
\renewcommand{\algorithmicendloop}{\algorithmicend}

\newcommand{\cst}[1]{\boldsymbol{\mathit{#1}}}
\newcommand{\culpritcst}{\cst{SpC}}
\newcommand{\newcst}{\cst{NewC}}
\newcommand{\culpritblk}{\mathit{SpB}}
\newcommand{\splittedblk}{\mathit{RfnB}}
\newcommand{\newblk}{\mathit{NewB}}
\newcommand{\newbottomstates}{\mathit{NewBott}}
\newcommand{\nottau}{{\setminus\hspace*{-.4em}\tau}}

\newcommand{\shortto}{\mathord{\tikz[baseline]{\clip(0em,-0.3ex) rectangle (0.5em,1.8ex); \node[draw=none,inner sep=0pt,anchor=base](0em,0ex){$\to$};}}}
\newcommand{\bigo}[2][]{\mathcal{O}#1\left(#2\right)}
\newcommand{\bigOmega}[2][]{\Omega#1\left(#2\right)}

\newcommand{\bluesquare}{\raisebox{-0.4ex}[0pt][0pt]{\tikz \draw (1ex,1ex) node[minimum size=1.8ex,inner sep=0pt,rectangle,draw=blue,fill=blue!50]{};}\hspace*{0.1em}}
\newcommand{\orangecirc}{\raisebox{-0.4ex}[0pt][0pt]{\tikz \draw (1ex,1ex) node[minimum size=2ex,inner sep=0pt,circle,draw=orange,fill=orange!50]{};}\hspace*{0.1em}}
\newcommand{\blackdiamond}{\raisebox{-0.4ex}[0pt][0pt]{\tikz \draw (1ex,1ex) node[minimum size=2ex,inner sep=0pt,diamond,draw=black,fill=black!50]{};}\hspace*{0.1em}}
\newcommand{\blackpentagon}{\raisebox{-0.4ex}[0pt][0pt]{\tikz \draw (1ex,1ex) node[minimum size=2ex,inner sep=0pt,regular polygon,draw=black,fill=black!50]{};}\hspace*{0.1em}}
\newcommand{\redtriangle}{\raisebox{-0.4ex}[0pt][0pt]{\tikz \draw (1ex,1ex) node[minimum size=2ex,inner sep=0pt,isosceles triangle,rotate=90,draw=red,fill=red!50]{};}\hspace*{0.1em}}

\title{Stuttering equivalence is too slow!}
\author{	\parbox[t]{4cm}{\centering David N. Jansen\textsuperscript{\dag} \\ \normalsize \href{mailto:dnjansen@cs.ru.nl}{dnjansen@cs.ru.nl}}
\and
\parbox[t]{4cm}{\centering Jeroen J. A. Keiren\textsuperscript{*,\dag} \\ \normalsize \href{mailto:Jeroen.Keiren@ou.nl}{Jeroen.Keiren@ou.nl}}
}

\date{\begin{minipage}{\textwidth}\centering
	\parbox[t]{12.8cm}{\raggedright\footnotesize
		\textsuperscript{*}Open University of the Netherlands, School of Computer Science, Netherlands \\
		\textsuperscript{\dag}Radboud Universiteit, Institute for Computing and Information Sciences, Nijmegen, Netherlands
	}
\end{minipage}
}

\begin{document}

\maketitle

\begin{abstract}
Groote and Wijs recently described an algorithm for deciding stuttering equivalence and branching bisimulation equivalence, acclaimed to run in $\bigo{m \log n}$ time.
Unfortunately, the algorithm does not always meet the acclaimed running time.
In this paper, we present two counterexamples where the algorithms uses $\bigOmega{md}$ time.
A third example shows that the correction is not trivial.
In order to analyse the problem we present pseudocode of the algorithm,
and indicate the time that can be spent on each part of the algorithm
in order to meet the desired bound.
We also propose fixes to the algorithm such that it indeed runs in $\bigo{m \log n}$ time.
\end{abstract}

\begin{algorithm}[t]
\caption{The main algorithm for stuttering equivalence. Closely follows \cite{GrooteW16}\label{algo:feb11}}
\begin{algorithmic}[1]
\STATE	Initialise all temporary data.
	\rightbrace{1}{$\bigo{m \log n}$}
\WHILE{there is a nontrivial constellation}
	\STATE	Choose a nontrivial constellation $\culpritcst$
		\\ and a splitter $\culpritblk \subset \culpritcst$ that is small {\small (i.\,e.\@ $\lvert \culpritblk \rvert \leq \frac{1}{2} \lvert \culpritcst \rvert$).}
		\rightbrace[-1]{4}{$\bigo{1}$ per splitter $\culpritblk$}
	\STATE	Create a new constellation $\newcst$ \\ and move $\culpritblk$ from $\culpritcst$ to $\newcst$.
	\FORALL[Find predecessors of $\culpritblk$]{$s \in \culpritblk$}
		\FORALL{$s' \in \mathit{in}(s) \setminus \culpritblk$}
			\STATE	Mark the block of $s'$ as refinable.
				\rightbrace[-2]{7}{$\bigo{\lvert \mathit{in}(\culpritblk) \rvert}$}
			\STATE	\label{alg:markstates}
				Mark $s'$ as predecessor of $\culpritblk$.
			\STATE	\label{alg:registertransition}
				Register that $s' \to s$ goes to $\newcst$ (instead of $\culpritcst$).
		\ENDFOR
	\ENDFOR
	\STATE	Prepare $\culpritblk$ to be refined {\small (i.\,e.\@, mark states, register transitions).}
		\rightbrace{1}{$\bigo{\lvert \mathit{out}(\culpritblk) \rvert}$}
	\BEGIN[Stabilise the partition again:]
		\FORALL{refinable blocks $\splittedblk$}
			\STATE	$\mathit{Result} := \textsc{TrySplit}(\splittedblk, \newcst, \text{marked states} \in \splittedblk,\mbox{}$
				\\$\phantom{\mathit{Result} := \textsc{TrySplit}(}$%
				unmarked bottom states $\in \splittedblk)$
				\label{alg:splitNew}
			\STATE	\label{alg:splitBminusC}
				$\textsc{TrySplit}'(\mathit{Result}, \culpritcst,
				\text{states} \in \mathit{Result} \text{ with a transition to } \culpritcst,
				\mbox{}$\\$\phantom{\textsc{TrySplit}'(}
				\text{bottom states} \in \mathit{Result} \text{ without transition to } \culpritcst)$
			\STATE	$\textsc{PostprocessNewBottom}()$
				\label{alg:call-processnewbottomstates}
		\ENDFOR
	\END
	\STATE	Destroy all temporary data {\small (i.\,e.\@, markings of states and blocks).}
		\rightbrace{1}{$\bigo{\lvert \mathit{in}(\culpritblk) \rvert {+} \lvert \mathit{out}(\culpritblk) \rvert}$}
		\mbox{}
\ENDWHILE
\end{algorithmic}
\end{algorithm}

\begin{algorithm}
\caption{Refine a block into red and blue states, called in Line~\ref{alg:splitNew}.
Slightly improved\label{alg:trysplit}}
\algsetup{indent=1em}
\begin{algorithmic}[1]
\STATE	\textbf{function} $\textsc{TrySplit}(\splittedblk, \culpritcst, \mathit{Red}, \mathit{Blue})$
\STATE	\COMMENT{Try to refine block $\splittedblk$,
	depending on whether states have transitions to the splitter con\-stel\-la\-tion $\culpritcst$.
	$\mathit{Red}$ contains all states in $\splittedblk$ with a strong transition to $\culpritcst$,
	and $\mathit{Blue}$ contains all bottom states in $\splittedblk$ without transition to $\culpritcst$.}
\BEGIN[Spend the same amount of work on either process:]
	\begin{tabular}{@{}l@{\,}||l}
	\parbox[t]{5.65cm}{
		\vspace*{-0.65\baselineskip}
		\STATE	\textbf{whenever} $\lvert \mathit{Blue} \rvert > \frac{1}{2}\lvert \splittedblk \rvert$ \algorithmicthen
		\STATE	\mbox{\hspace*{\algorithmicindent}}Stop this process.
		\STATE	\textbf{end whenever}
		\WHILE{$\mathit{Blue}$ contains \\ \mbox{}\hfill unvisited states} \label{alg:trysplit:while}
			\label{alg:trysplit:whilebodybegin}
			\STATE	Choose an unvisited $s \in \mathit{Blue}$.
			\STATE	Mark $s$ as visited.
			\FORALL{$s' \in \mathit{in}_\tau(s) \setminus \mathit{Red}$}
				\label{alg:trysplit:forallpred}
				\IF{$\mathit{notblue}(s')$ undefined}
					\STATE	\label{alg:trysplit:transcount}
						$\mathit{notblue}(s') := \lvert \mathit{out}_\tau(s') \rvert$
				\ENDIF
				\STATE	$\mathit{notblue}(s') \mathbin{:=} \mathit{notblue}(s') {-} 1$
				\IF{$\mathit{notblue}(s') = 0$}
					\STATE	$\mathit{Blue} := \mathit{Blue} \cup \{ s' \}$
				\ENDIF
			\ENDFOR
		\ENDWHILE \label{alg:trysplit:endwhile}
		\STATE	Stop the other process.
		\STATE	\label{alg:trysplit:movebluetonew}
			Move $\mathit{Blue}$ to a new block $\newblk$.
		\STATE	Destroy all temporary data.
		\FORALL{$s \in \newblk$}
			\label{alg:searchnewbottomstates:start}
			\FORALL{$s' \in \mathit{in}_\tau(s) \setminus \newblk$}
				\STATE	$s' \to s$ is no longer inert.
				\IF{$\lvert \mathit{out}_\tau(s') \rvert = 0$}
					\STATE	$s'$ is a new bottom state.
						\label{alg:newbottomstatefound-blue}
				\ENDIF
				\label{alg:newbottomstatefound-red}
			\ENDFOR
		\ENDFOR
			\label{alg:searchnewbottomstates:end}
		\STATE	$\mathit{Result} := \splittedblk$
		\vspace*{0.18\baselineskip}
	} &
	\parbox[t]{8.63cm}{
		\vspace*{-0.65\baselineskip}
		\item[]	\textbf{whenever} $\lvert \mathit{Red} \rvert {>} \frac{1}{2}\lvert \splittedblk \rvert$ \algorithmicthen
			\rightbrace{3}{\parbox{2.65cm}{$\bigo{1}$ per assign\-ment to $\mathit{Blue}$ or $\mathit{Red}$, respectively.}}
		\item[]	\mbox{\hspace*{\algorithmicindent}}Stop this process.
		\item[]	\textbf{end whenever}
		\item[]	\algorithmicwhile{} $\mathit{Red}$ contains
			\rightbrace{14}{$\bigo{\lvert \mathit{in}_\tau(\newblk) \rvert}$}
		\item[]	\mbox{}\hfill unvisited states \algorithmicdo\mbox{\hspace*{3.8cm}}
		\item[]	\mbox{\hspace*{\algorithmicindent}}Choose an unvisited $s \in \mathit{Red}$.
		\item[]	\mbox{\hspace*{\algorithmicindent}}Mark $s$ as visited.
		\item[]	\mbox{\hspace*{\algorithmicindent}}\algorithmicforall{} $s' \in \mathit{in}_\tau(s)$ \algorithmicdo
		\item[]	\mbox{}
		\item[]	\mbox{}
		\item[]	\mbox{}
		\item[]	\mbox{}
		\item[]	\mbox{}
		\item[]	\mbox{\hspace*{2\algorithmicindent}}$\mathit{Red} := \mathit{Red} \cup \{ s' \}$
		\item[]	\mbox{}
		\item[]	\mbox{\hspace*{\algorithmicindent}}\algorithmicendfor
		\item[]	\algorithmicendwhile
		\item[]	Stop the other process.
			\rightbrace{2}{$\bigo{\lvert \mathit{out}(\newblk) \rvert}$}
		\item[]	Move\,$\mathit{Red}$\,to\,a\,new\,block\,$\newblk$.
		\item[]	Destroy all temporary data.
			\rightbrace{1}{as Lines \ref{alg:trysplit:while}--\ref{alg:trysplit:endwhile}}
		\item[]	\algorithmicforall{} $s \in \newblk$ \algorithmicdo
			\rightbrace{8}{$\begin{gathered}
				\bigo{\lvert \mathit{in}_\tau(\newblk) \rvert} \hfill \\[-2pt]
				\text{or} \hfill \\
				\bigo{\lvert \mathit{out}_\tau(\newblk) \rvert}
			\end{gathered}$}
		\item[]	\mbox{\hspace*{\algorithmicindent}}\algorithmicforall\,$s' \in \mathit{out}_\tau(s) {\setminus} \newblk$\,\algorithmicdo
		\item[]	\mbox{\hspace*{2\algorithmicindent}}$s \to s'$ is no longer inert.
		\item[]	\mbox{\hspace*{\algorithmicindent}}\algorithmicendfor
		\item[]	\mbox{\hspace*{\algorithmicindent}}\algorithmicif{} $\lvert \mathit{out}_\tau(s) \rvert = 0$ \algorithmicthen
		\item[]	\mbox{\hspace*{2\algorithmicindent}}$s$ is a new bottom state.
		\item[]	\mbox{\hspace*{\algorithmicindent}}\algorithmicendif
		\item[]	\algorithmicendfor
		\item[]	$\mathit{Result} := \newblk$
		\vspace*{0.18\baselineskip}
	}
	\end{tabular}
\END
\STATE	\algorithmicreturn{} $\mathit{Result}$
\end{algorithmic}
\end{algorithm}

\section{Introduction}
It has long been an open problem whether the algorithm by Groote and Vaandrager \cite{GrooteV90} for computing stuttering equivalence \cite{browne_characterizing_1988} and branching bisimulation \cite{van_glabbeek_branching_1996} was optimal.
Recently, Groote and Wijs \cite{GrooteW16,GrooteW16tacas} presented an improvement.
They describe an algorithm for deciding stuttering equivalence in time $\bigo{m \log n}$ and space $\bigo{m}$, where $n$ is the number of states, and $m$ the number of transitions of the Kripke structure at hand, with $m \geq n$. This is an improvement over the previous running time of $\bigo{mn}$.

Unfortunately, the algorithm \cite{GrooteW16} falls short of the stated goal. In this paper we introduce two counterexamples where the algorithm will use more time than $\bigo{m \log n}$, namely $\bigOmega{m d}$, where $d$ is the maximal outdegree of a state in the Kripke structure.

Since the original description of the algorithm relies heavily on auxiliary data structures and pointers in order to ensure that all information is available quickly when needed, without making the data structures any bigger than strictly necessary, the problem with the algorithm is hard to detect in this original description. We therefore first present our understanding of the algorithm as presented in \cite{GrooteW16} by giving a high-level description in pseudocode,
which leaves out as much of the detailed data structures as possible.
This also allows us to assign time budgets to its parts,
that are satisfied locally
and that together allow us to determine the overall time bound.
Subsequently, we identify two problems in the original algorithm, by giving counterexamples that lead to running times higher than the desired bound of $\bigo{m \log n}$. For each of these problems, we indicate how to fix the algorithm. When fixing the second problem, a further complication arises that needs to be resolved in order to meet the desired bound.
Yet, ultimately we can confirm the main result of \cite{GrooteW16} that stuttering equivalence and branching bisimulation can be computed in $\bigo{m \log n}$ time and $\bigo{m}$ space.

Throughout this paper we assume the reader is familiar with the definitions of Kripke structure and (divergence-blind) stuttering equivalence and with the auxiliary notions introduced in~\cite{GrooteW16}. The note is best read while having a copy of \cite{GrooteW16} within reach for reference. We focus our analysis on deciding stuttering equivalence for Kripke structures. The results carry over directly to the computation of branching bisimulation.

\section{Pseudocode for Groote/Wijs 2016}

We first present the main part of the algorithm from \cite{GrooteW16} as we understand it in terms of pseudocode, while separating out a routine $\textsc{TrySplit}$
that tries to split a refinable block
into states that can reach the splitter (called red states)
and those that cannot (called blue states).
There is a small difference between the calls to $\textsc{TrySplit}$
in lines~\ref{alg:splitNew} and \ref{alg:splitBminusC},
which we will explain later.
The high-level structure of the algorithm is presented in Algorithm~\ref{algo:feb11}.
It maintains the invariant:
\begin{invariant}
	\label{inv:stablewrtcst}
	The current blocks are stable with respect to the constellations,
	i.\,e.\@ all states in a block can reach the same constellations through a (weak) transition.
\end{invariant}
A nontrivial constellation, i.\,e.\@ a constellation containing multiple blocks,
indicates that the current blocks are not yet stable with respect to themselves.
The main loop separates a block $\culpritblk$
from a nontrivial constellation $\culpritcst$,
moving it to a new constellation $\newcst$,
and then restores the invariant by refining blocks with respect to these new constellations, if needed.
The latter
is done in $\textsc{TrySplit}$.

In the algorithms we assigned a time budget to some steps
to facilitate the analysis of
the algorithms' complexity.
We generally use the abbreviations $\lvert \mathit{in}(B) \rvert = \sum_{s \in B} \max\{1, \lvert \mathit{in}(s) \rvert \}$
and $\lvert \mathit{out}(B) \rvert = \sum_{s \in B} \max\{1, \lvert \mathit{out}(s) \rvert \}$.
Similarly, $\mathit{in}_\tau(s)$ are the \emph{inert} incoming transitions
and $\mathit{out}_\tau(s)$ the \emph{inert} outgoing transitions of $s$.

\subsection{Splitting blocks}

The most important new idea from \cite{GrooteW16} is used in the step
when a block is actually being refined.
They start to find both the red and the blue states,
spending the same amount of work on either part,
until it becomes clear which one is the smaller.
In other words, they use the idea ``Process the smaller half'' not only when looking for splitters,
but also when refining blocks.
The work spent on the refinement can then be bounded:
state $s$ is involved in such a refinement at most $\bigo{\log n}$ times.
Upon every such refinement,
at most $\bigo{\lvert \mathit{in}(s) \rvert + \lvert \mathit{out}(s) \rvert}$ time
is spent for state $s$.
So, overall $\bigo{(\lvert \mathit{in}(s) \rvert + \lvert \mathit{out}(s) \rvert) \log n}$ time
is spent on state $s$.
Summing over all states then gives the desired time complexity $\bigo{m \log n}$.

Note that this may require detailed bookkeeping of the amount of work.
One may balance the work by using an auxiliary variable,
which stores the amount of work done on the red states minus the amount of work done on the blue states.
Every time a state or transition is checked
(i.\,e.\@ every time the loop in Lines~\ref{alg:trysplit:whilebodybegin} or \ref{alg:trysplit:forallpred}
is entered),
the balance is increased or reduced by 1.

We deviate from \cite{GrooteW16} slightly: \emph{ibid.} uses a priority queue to keep track of
$\mathit{notblue}$, but actually nothing queue-like is needed,
as the order of states is irrelevant for the correctness, time and memory bounds.
Note that the data structure should allow to test for membership
in time $\bigo{1}$.\footnote{Note that priority queues typically have longer access times.}
They use the priority only to store the value $\mathit{notblue}(s')$
and check whether this value is defined by a test for membership in the priority queue.
We propose instead to set $\mathit{notblue}(s')$ to some special value (e.\,g.\@ 0)
to indicate that it was not yet calculated.
We also need a list or set of all states whose $\mathit{notblue}(s')$ is defined,
to destroy the temporary data later.

\begin{algorithm}[tbp]
\caption{Refine as required by new bottom states, called in Line~\ref{alg:call-processnewbottomstates}\label{algo:newbottomsplit}}
\begin{minipage}{\textwidth}
\begin{algorithmic}[1]
\STATE	\textbf{function} $\textsc{PostprocessNewBottom}()$
	\rightbrace[1]{3}{\parbox{3.7cm}{$\bigo{\lvert \mathit{out}(s) \rvert}$ for some old bottom state $s \in \hat{B}$}}
\FORALL{constellations $\cst{C}$ reachable from $\hat{B}$}
	\STATE	Register that the pair $(\hat{B}, \cst{C})$ needs postprocessing.
		\label{alg:registerbottomreach}
\ENDFOR
\WHILE{there is a pair $(\hat{B}, \cst{C})$ that needs postprocessing}
	\STATE	Choose a pair $(\hat{B}, \cst{C})$ that needs postprocessing.
		\rightbrace{3}{$\bigo{1}$ per pair $(\hat{B}, \cst{C})$}
	\STATE	Delete $(\hat{B}, \cst{C})$ from the pairs that need postprocessing.
	\IF{not all new bottom states can reach $\cst{C}$}
		\STATE	$\textsc{TrySplit}'(\hat{B}, \cst{C},
			\text{states} \in \hat{B} \text{ with a transition to } \cst{C},
			\mbox{}$\\$\phantom{\textsc{TrySplit}'(}
			\text{new bottom states} \in \hat{B} \text{ without transition to } \cst{C})$
			\label{alg:trysplit:newbottom}
			\hspace*{-1cm}
			\rightbrace[1]{5}{$\bigo{\lvert \mathit{out}(\newblk) \rvert}$}
		\FORALL{constellations $\cst{C'}$ that $\newblk$ can reach\footnote{$\newblk$
			is the new block created in $\textsc{TrySplit}'$, Line~\ref{alg:trysplit:newbottom}.}}
			\label{alg:newattention-begin}
			\IF{$(\hat{B}, \cst{C'})$ still needs postprocessing}
				\STATE	Register that $(\newblk, \cst{C'})$ needs postprocessing.
			\ENDIF
		\ENDFOR	\label{alg:newattention-end}
	\ENDIF
\ENDWHILE
\STATE	Destroy all temporary data.
\RETURN
\end{algorithmic}
\end{minipage}
\end{algorithm}

\subsection{New bottom states}

While refining a block, it may happen that some states become bottom states
because all their inert transitions become transitions from a red state to a blue state
and therefore are no longer inert.
We have to single out these new bottom states
because the algorithm treats them differently from the non-bottom states.
Here, we also added a slight improvement
(Lines~\ref{alg:searchnewbottomstates:start}--\ref{alg:searchnewbottomstates:end}):
we only look for new bottom states in $\mathit{Red}$.

Additionally, it may happen
that a new bottom state can no longer reach all the constellations
that were reachable from the original bottom states.
To repair Invariant~\ref{inv:stablewrtcst},
we may have to split some new bottom states off the block.
This further splitting itself is shown in Algorithm~\ref{algo:newbottomsplit}.
The basic idea is to check, for each constellation that is reachable from some new bottom state,
whether the block has to be split.
Of course, as soon as a block is split, both parts have to be checked for the remaining reachable constellations.
The algorithm in \cite{GrooteW16} constructs, for each block $\hat{B}$ and constellation $\cst{C}$,
a list $S_{\cst{C}} \subseteq \hat{B}$ of states that can reach $\cst{C}$ (see Line~\ref{alg:registerbottomreach}).
These lists are used in Line~\ref{alg:trysplit:newbottom} to decide which states are blue,
namely the new bottom states that are \emph{not} in $S_{\cst{C}}$.
The time budget $\bigo{\lvert \mathit{out}(\newbottomstates) \rvert}$ can be met
if the list $S_{\cst{C}}$ follows the same order as the list $\newbottomstates$.

Algorithm~\ref{algo:newbottomsplit} generally follows \cite{GrooteW16},
except in Lines~\ref{alg:newattention-begin}--\ref{alg:newattention-end},
where we tried to find a formulation that fits in the time budget.

\section{Counterexamples to time complexity}

The algorithm contains two problems that cause it to be too slow.
First, the second call to $\textsc{TrySplit}$, here referred to as $\textsc{TrySplit}'$, takes too long.
We give a counterexample, and improve $\textsc{TrySplit}'$ to improve the complexity to the required bound.
Also the postprocessing of new bottom states as carried out in \cite{GrooteW16} is too slow.
In section~\ref{sec:postprocess_too_slow} we analyse the problem in the original algorithm;
the proposed improvement has already been incorporated in Algorithm~\ref{algo:newbottomsplit}.

\subsection{$\text{\sc TrySplit}'$ is too slow}

In the call to $\textsc{TrySplit}'$ (in Line~\ref{alg:splitBminusC}),
the initial set of red states is given implicitly,
through a list of transitions.
As a consequence, the test whether the potentially blue state $s'$ has a non-inert transition to $\culpritcst$
(this is why we require $s' \not\in \mathit{Red}$ in Line~\ref{alg:trysplit:forallpred}$\ell$)
in the variant $\textsc{TrySplit}'$
is executed in a different way compared to the one in \textsc{TrySplit}.
Groote and Wijs \cite{GrooteW16} add this test just before Line~\ref{alg:trysplit:transcount}$\ell$.
If $s'$ is marked (i.\,e.\@ it has a transition to $\culpritblk$),
they can access one of their many auxiliary variables,
but otherwise,
``it can be checked by walking over the transitions $s' \to s'' \in s'.T_\mathit{tgt}$''
(obviously, this is meant instead of the original ``$\ldots \in s.T_\mathit{tgt}$''
-- see Section 5.3, item 1.(b).ii.A.second bullet.first dash of \cite{GrooteW16}).
So they execute a loop to verify
$\mathit{out}_\nottau(s') \cap \culpritcst = \emptyset$, namely:

\vspace{-0.5\baselineskip}

\begin{center}
\begin{minipage}{0.6\textwidth}
\begin{algorithmic}
	\FORALL{$s'' \in \mathit{out}(s')$}
		\STATE	\algorithmicif{} $s'' \in \culpritcst \setminus \splittedblk$ \algorithmicthen{}
			\textbf{continue} to Line~\ref{alg:trysplit:forallpred}$\ell$
	\ENDFOR	\phantom{g}
\end{algorithmic}
\end{minipage}
\end{center}

\vspace{-0.4\baselineskip}

This loop makes their algorithm slower than promised:
the test uses time $\bigo{\lvert \mathit{out}(s') \rvert}$,
but should take at most $\bigo{1}$.
Our first counterexample illustrates this time budget overrun.

\begin{figure}[!ht]
\caption{$\textsc{TrySplit}'$ is too slow.\label{fig:counterexample}}
\begin{center}
\begin{tikzpicture}[>=Stealth]
	\fill[gray!5]
		(1.9,6.4) rectangle (4.5,8.6)
		(5.9,6.4) rectangle (10.9,8.6);
	\draw[very thin]
		(1.9,8.6) node[below right,inner sep=1mm]{$\culpritcst$} rectangle (10.9,6.4);
	\draw[very thin,fill=gray!5]
		(2.95,0)node[above right]{$\cst{E}_1$}	rectangle (5.55,2.2)
		(5.85,0)node[above right]{$\cst{E}_2$}	rectangle (8.45,2.2)
		(9.45,0)node[above right]{$\cst{E}_d$}	rectangle (12.05,2.2)
		(1.9,6.1)node[below right]{$\cst{D}$}	rectangle (13.1,2.5);
	\draw	(7.5,-0.1)	node[below]{$d$ constellations};

	\fill[fill=gray!15,rounded corners=8mm]	
		(2.2,2.8) rectangle (6.4,5.8)
		(6.7,2.8) rectangle (12.8,5.8);
	\draw[thick,rounded corners=8mm]	
		(2.2,5.8) rectangle (12.8,2.8)	node[above left,inner sep=1.7mm+2pt]{$\splittedblk$};
	\draw[thick,fill=gray!15,rounded corners=8mm]	
		(2.2,8.3) node[below right,inner sep=1.7mm+2pt]{$\culpritblk$}	rectangle (4.2,6.7)
		(8.6,6.7) rectangle (10.6,8.3)

		(3.25,0.3) node[above right,inner sep=1.7mm+2pt]{$E_1$}	rectangle (5.25,1.9)
		(6.15,0.3) node[above right,inner sep=1.7mm+2pt]{$E_2$}	rectangle (8.15,1.9)
		(9.75,0.3) node[above right,inner sep=1.7mm+2pt]{$E_d$}	rectangle (11.75,1.9);

	\node[minimum size=3mm,inner sep=0pt,diamond,draw,fill=black!50] at (4.25,1.1) (bot1){};
	\node[minimum size=3mm,inner sep=0pt,diamond,draw,fill=black!50] at (7.15,1.1) (bot2){};
	\node[inner sep=1pt] at (8.95,1.1) {$\ldots$};
	\node[minimum size=3mm,inner sep=0pt,diamond,draw,fill=black!50] at (10.75,1.1) (botd){};
	\draw[->]	(7.5,3.6) to (bot1);
	\draw[->]	(7.5,3.6) to (bot2);
	\draw[->]	(7.5,3.6) to (botd);
	\draw[->]	(3,5) .. controls (3,2) and (3.2,1.1) .. (bot1);
	\draw[->]	(3,5) .. controls (3,2) and (5,1.1) .. (bot2);
	\draw[->]	(3,5) .. controls (3,2) and (8,1.1) .. (botd);

	\node[minimum size=3mm,inner sep=0pt,circle,draw=orange,fill=orange!50] at (7.5,3.6) (r){};
	\node[minimum size=3mm,inner sep=0pt,regular polygon,draw=black,fill=black!50] at (3.2,7.5) (bluegoal){};
	\node[minimum size=3mm,inner sep=0pt,regular polygon,draw=black,fill=black!50] at (9.6,7.5) (redgoal){};
	\draw[->]	(3,5) to (bluegoal);
	\draw[->]	(9.4,5) to (r);

	\node[minimum size=2.8mm,inner sep=0pt,rectangle,draw=blue,fill=blue!50,text=white] at (3,5) (blue1){\footnotesize m};
	\draw[->]	(4,5) to (blue1);
	\node[minimum size=2.8mm,inner sep=0pt,rectangle,draw=blue,fill=blue!50] at (4,5)	(blue2){};
	\node[inner sep=1pt] at	(4.8,5) (bluedots){$\ldots$};
	\draw[->]	(bluedots) to (blue2);
	\draw[->]	(5.6,5) to (bluedots);
	\node[minimum size=2.8mm,inner sep=0pt,rectangle,draw=blue,fill=blue!50] at (5.6,5) (bluek){};
	\node at (4.3,5.5){$k$ states};

	\node[minimum size=3mm,inner sep=0pt,isosceles triangle,rotate=90,draw=red,fill=red!50] at (9.4,5) (red1){};
	\draw[->]	(10.4,5) to (red1);
	\node[minimum size=3mm,inner sep=0pt,isosceles triangle,rotate=90,draw=red,fill=red!50] at (10.4,5) (red2){};
	\node[inner sep=1pt] at (11.2,5) (reddots){$\ldots$};
	\draw[->]	(reddots) to (red2);
	\draw[->]	(12,5) to (reddots);
	\node[minimum size=3mm,inner sep=0pt,isosceles triangle,rotate=90,draw=red,fill=red!50] at (12,5)	(redk){};
	\node at (10.7,5.5){$k$ states};
	\draw[->]	(r) to (blue1);
	\draw[->]	(r) to (blue2);
	\draw[->]	(r) to (bluek);
	\draw[->]	(r) to (redgoal);
\end{tikzpicture}
\end{center}
\end{figure}

Assume that the partition shown in Figure~\ref{fig:counterexample} has been reached.
Then, we refine $\culpritcst$:
we select $\culpritblk$, find its weak predecessors
(the whole block $\splittedblk$, so nothing is refined)
and the weak predecessors of $\culpritcst \setminus \culpritblk$
(the right half of block $\splittedblk$).
Note that we do the latter
without walking over the states in $\culpritcst \setminus \culpritblk$:
it is ok to spend $\bigo{\lvert \mathit{in}(\culpritblk) \rvert}$ time here.
We also save time by calculating the \emph{complement} of the weak predecessors,
i.\,e.\@ the \bluesquare-states in $\newblk$,
because it is smaller than $\splittedblk \setminus \newblk$:
we are allowed to spend an additional $\bigo{\lvert \mathit{in}(\newblk) \rvert + \lvert \mathit{out}(\newblk) \rvert}$ time on this task.

As the algorithm looks through the predecessors of all \bluesquare-states,
it considers the \orangecirc-state for inclusion in $\newblk$.
This happens $k$ times
(once for each transition to a \bluesquare-state).
The cited passage then requires
that we check each time
whether some immediate successor of a \orangecirc-state is in $\culpritcst \setminus \culpritblk$.
If we check the transitions to the \blackdiamond-states before the \blackpentagon-state,
we spend $\Omega(d)$ time,
and in the end we find that the \orangecirc-state should not be considered further.\footnote{In
	particular, \cite{GrooteW16} does not define $\mathit{notblue}(\orangecirc)$.
	An ad-hoc solution would be to set it to some value $> d$,
	but that does not help if there are multiple copies of \orangecirc.}
The problem is that we spend (much) time on a transition to $\newblk$
from a state that is \emph{possibly} in $\newblk$
but in the end not \emph{actually} in $\newblk$.
The part of the algorithm in Section 5.3, 1.(b).ii.A.second bullet, 
is only allowed to spend $\bigo{1}$ time on each such transition.

Overall, the number of states and transitions both are $\bigo{k+d}$.
So, the algorithm is allowed to spend $\bigo{(k+d) \log (k+d)}$ time,
but it spends $\bigOmega{k d}$ time.
Variants of this Kripke structure with several copies of the \orangecirc-state
show that the checks can cost $\bigo{md}$ time.

Note that we are not allowed to concentrate on the red states
(the weak predecessors of $\culpritcst \setminus \culpritblk$, the \redtriangle-states and \orangecirc-state)
themselves instead of the complement,
as this set is larger.

\begin{lemma}
	\label{lem:timebound1}
	Refining in Lines~\ref{alg:splitBminusC} and \ref{alg:trysplit:newbottom}
	as described in \cite{GrooteW16}
	has a worst-case time complexity of $\bigOmega{m d}$.
\end{lemma}

\begin{algorithm}[tbp]
\caption{Refine a block w.\,r.\,t.\@ $\culpritcst \setminus \culpritblk$ (corrected), called in Line~\ref{alg:splitBminusC}\label{algo:trysplitaccent}}
\algsetup{indent=1em}
\begin{algorithmic}[1]
\STATE	\textbf{function} $\textsc{TrySplit}'(\splittedblk, \culpritcst, \mathit{FromRed}, \mathit{MaybeBlue}, \mathit{isBlueTest})$
\STATE	\COMMENT{Try to refine block $\splittedblk$,
	depending on whether states have transitions to the splitter con\-stel\-la\-tion $\culpritcst$.
	$\mathit{FromRed}$ contains all transitions from $\splittedblk$ to $\culpritcst$,
	$\mathit{MaybeBlue}$ contains all bottom states that may be initially blue states,
	$\mathit{isBlueTest}$ is a predicate that determines
	whether a candidate in $\mathit{MaybeBlue}$ is definitely blue.}
\BEGIN[Spend the same amount of work on either process:]
	\begin{tabular}{@{}l@{\,}||l}
	\parbox[t]{5.65cm}{
		\vspace*{-0.65\baselineskip}
		\STATE	$\mathit{Blue} := \emptyset$
		\STATE	\textbf{whenever} $\lvert \mathit{Blue} \rvert > \frac{1}{2}\lvert \splittedblk \rvert$ \algorithmicthen
		\STATE	\mbox{\hspace*{\algorithmicindent}}Stop this process.
		\STATE	\textbf{end whenever}
		\WHILE{$\mathit{Blue}$ or $\mathit{MaybeBlue}$ contain \\\mbox{}\hfill unvisited states}
			\label{alg:while:red}
			\STATE	Choose an unvisited $s \in \mathit{Blue}$ \\ \mbox{}\hfill or $s \in \mathit{MaybeBlue}$.
				 \label{alg:loopbody:while:red}
			\STATE	Mark $s$ as visited.
			\IF{$s \not\in \mathit{Blue}$}
				\IF{$\neg isBlueTest(s)$}
					\label{alg:trysplitprime:bluetest}
					\STATE	\textbf{continue} to Line~\ref{alg:while:red}$\ell$
				\ENDIF
				\STATE	$Blue := Blue \cup \{ s \}$
			\ENDIF
			\STATE	\mbox{}
			\FORALL{$s' \in \mathit{in}_\tau(s) \setminus \mathit{Red}$}
				\IF{$\mathit{notblue}(s')$ undefined} \label{alg:loopbody:forallred}
					\STATE	$\mathit{notblue}(s') := \lvert \mathit{out}_\tau(s') \rvert$
				\ENDIF
				\STATE	$\mathit{notblue}(s') \mathbin{:=} \mathit{notblue}(s') {-} 1$
				\IF{$\mathit{notblue}(s') = 0$}
					\IF{$\mathit{out}_\nottau(s') \cap \culpritcst = \emptyset$} \label{alg:slowtestred}
						\STATE	$\mathit{Blue} := \mathit{Blue} \cup \{ s' \}$
					\ENDIF
				\ENDIF
			\ENDFOR
		\ENDWHILE \label{alg:endwhile:red}
		\STATE	Stop the other process.
		\STATE	Move $\mathit{Blue}$ to a new block $\newblk$.
			\label{alg:trysplitprime:movetonewblk}
		\STATE	Destroy all temporary data.
	} &
	\parbox[t]{8.63cm}{
		\vspace*{-0.65\baselineskip}
		\item[]	$\mathit{Red} := \emptyset$
			\rightbrace{1}{$\bigo{1}$}
		\item[]	\textbf{whenever} $\lvert \mathit{Red} \rvert {>} \frac{1}{2}\lvert \splittedblk \rvert$ \algorithmicthen
			\rightbrace{3}{\parbox{2.65cm}{$\bigo{1}$ per assign\-ment to $\mathit{Blue}$ or $\mathit{Red}$, respectively.}}
		\item[]	\mbox{\hspace*{\algorithmicindent}}Stop this process.
		\item[]	\textbf{end whenever}
		\item[]	\algorithmicwhile{} $\mathit{Red}$ or $\mathit{FromRed}$ contain
			\rightbrace{24}{$\begin{gathered}
				\bigo[\!]{\!\!\!\begin{gathered}
						\lvert \mathit{in}_\tau(\newblk) \rvert + \\
						\hspace{-.1em}\lvert \mathit{MaybeBlue} \rvert +\hspace{.1em} \\
						\lvert \mathit{out}(\newblk) \rvert + \\
						\!\lvert \mathit{out}(\newbottomstates) \rvert
					\end{gathered}\!\!\!} \\
				\text{and} \hfill \\
				\bigo[\!]{\!\begin{gathered}
						\lvert \mathit{in}_\tau(\newblk) \rvert + \\
						\!\lvert \mathit{FromRed} \rvert
					\end{gathered}\!} \hfill \\
				\mbox{} \\[-3pt] \mbox{}
			\end{gathered}$}
		\item[]	\mbox{}\hfill unvisited elements \algorithmicdo{}\mbox{\hspace*{3.8cm}}
		\item[]	\mbox{\hspace*{\algorithmicindent}}Choose an unvisited $s \in \mathit{Red}$
		\item[]	\mbox{}\hfill or $s \to s'' \in \mathit{FromRed}$.\mbox{\hspace*{3.8cm}}
		\item[]	\mbox{\hspace*{\algorithmicindent}}\algorithmicif{} $s \to s''$ is chosen \algorithmicthen
		\item[]	\mbox{\hspace*{2\algorithmicindent}}Mark $s \to s''$ as visited.
		\item[]	\mbox{\hspace*{2\algorithmicindent}}\algorithmicif{} $s$\,was\,visited\,earlier \algorithmicthen
		\item[]	\mbox{\hspace*{3\algorithmicindent}}\textbf{continue} to Line~\ref{alg:while:red}$r$
		\item[]	\mbox{\hspace*{2\algorithmicindent}}\algorithmicendif
		\item[]	\mbox{\hspace*{2\algorithmicindent}}$\mathit{Red} := \mathit{Red} \cup \{ s \}$
		\item[]	\mbox{\hspace*{\algorithmicindent}}\algorithmicendif
		\item[]	\mbox{\hspace*{\algorithmicindent}}Mark $s$ as visited.
		\item[]	\mbox{\hspace*{\algorithmicindent}}\algorithmicforall{} $s' \in \mathit{in}_\tau(s)$ \algorithmicdo
		\item[]	\mbox{}
		\item[]	\mbox{}
		\item[]	\mbox{}
		\item[]	\mbox{}
		\item[]	\mbox{}
		\item[]	\mbox{}
		\item[]	\mbox{\hspace*{2\algorithmicindent}}$\mathit{Red} := \mathit{Red} \cup \{ s' \}$
		\item[]	\mbox{}
		\item[]	\mbox{}
		\item[]	\mbox{\hspace*{\algorithmicindent}}\algorithmicendfor
		\item[]	\algorithmicendwhile
		\item[]	Stop the other process.
			\rightbrace{2}{$\bigo{\lvert \mathit{out}(\newblk) \rvert}$}
		\item[]	Move\,$\mathit{Red}$\,to\,a\,new\,block\,$\newblk$.
		\item[]	Destroy all temporary data.
			\rightbrace{1}{as Lines \ref{alg:while:red}--\ref{alg:endwhile:red}}
	}
	\end{tabular}
	\vspace*{-0.25\baselineskip}
	\STATE	Find new non-inert transitions and bottom states
		(as Lines~\ref{alg:searchnewbottomstates:start}--\ref{alg:searchnewbottomstates:end}).
		\label{alg:findnewbottom:start}
\END
\RETURN
\end{algorithmic}
\end{algorithm}

We tried (in vain) to find a recursive counterexample,
which should increase the time complexity to $\bigOmega{m d \log n}$,
but every of our ideas led to a counterexample with so many additional transitions
that it still fit the bound of Lemma~\ref{lem:timebound1}.

\subsection{A faster $\text{\sc TrySplit}'$}
\label{sec:faster-TrySplitprime}

We propose to solve this problem as follows:
\emph{Execute the slow test at the latest possible moment,}
namely immediately before a state is inserted in $\mathit{Blue}$.
This is shown in Algorithm~\ref{algo:trysplitaccent}.
Here, we also present the formal parameter list
according to the implicit representation of the red and blue states:
a set $\mathit{FromRed}$ of transitions from red states,
a set $\mathit{MaybeBlue}$ of possibly blue bottom states
with a predicate $\mathit{isBlueTest}$ that indicates
which bottom states are actually blue.
The overall time budget is still met:
If the red states are the smaller part,
then $\mathit{FromRed}$ is a subset of $\mathit{out}(\newblk)$.
If the blue states are the smaller part,
then states in $\mathit{MaybeBlue} \setminus \newblk$ are marked
(i.\,e.\@ they have a transition to $\culpritblk$, see Line~\ref{alg:markstates});
we are allowed to walk over them one more time.

When the test is executed in Line~\ref{alg:slowtestred}$\ell$,
all inert transitions of $s'$ point to blue states.
If $s'$ has a non-inert transition to $\culpritcst$,
it is actually a red state, and in particular, a red new bottom state.
(It may happen that we do not find all new bottom states here;
therefore, we still have to execute Line~\ref{alg:findnewbottom:start}$\ell$.)
As every state becomes a bottom state at most once,
we are allowed to spend time $\bigo{(\lvert \mathit{in}(s') \rvert + \lvert \mathit{out}(s') \rvert) \log n}$ then,
which is abundant.
If no such transition to $\culpritblk$ is found,
$s'$ is a blue state
and we have to account for the time differently.
It is $\bigo{\lvert \mathit{out}(s') \rvert}$ per unmarked blue state $s'$.
Every time $s'$ becomes an unmarked blue state, the test is executed exactly once,
which fits in the general bound per time that $s'$ is involved in a refinement.

\subsection{$\text{\sc PostprocessNewBottom}$ is too slow}
\label{sec:postprocess_too_slow}

In Algorithm~\ref{algo:newbottomsplit}, we already included an improvement
in Lines~\ref{alg:newattention-begin}--\ref{alg:newattention-end}.
If some block $\hat{B}$ is split here,
it should not take longer than $\bigo{\lvert \mathit{in}(\newblk) \rvert + \lvert \mathit{out}(\newblk) \rvert}$.
The original formulation did not take this into account;
it always walked over all lists $S_{\cst{C}}$ to separate them
into the part that belongs to $\newblk$ and the part that belongs to what remains in $\hat{B}$.
With our terminology, it did:

\vspace{-0.5\baselineskip}

\begin{center}
\begin{minipage}{0.8\textwidth}
\begin{algorithmic}
	\FORALL{constellations $\cst{C}$ such that $(\hat{B}, \cst{C})$ still needs postprocessing}
		\FORALL{new bottom states $s \in$ the original $\hat{B}$ with a transition to $\cst{C}$} 
			\IF{$s \in \newblk$}
				\STATE	Register that $(\newblk, \cst{C})$ needs postprocessing.
				\STATE	Move $s$ from $S_{\cst{C}}$ for $\hat{B}$ to the corresponding list for $\newblk$.
				\IF{$S_{\cst{C}}$ for $\hat{B}$ is empty}
					\STATE	Register that $(\hat{B}, \cst{C})$ no longer needs postprocessing.
				\ENDIF
			\ENDIF
		\ENDFOR
	\ENDFOR
\end{algorithmic}
\end{minipage}
\end{center}

\vspace{-0.4\baselineskip}

If $\newblk$ is much smaller than $\hat{B}$, a budget overrun may result
because the loop still spends (a little) time
for each state in $\hat{B} \setminus \newblk$ with a transition to $\cst{C}$.
This is illustrated in our second counterexample.

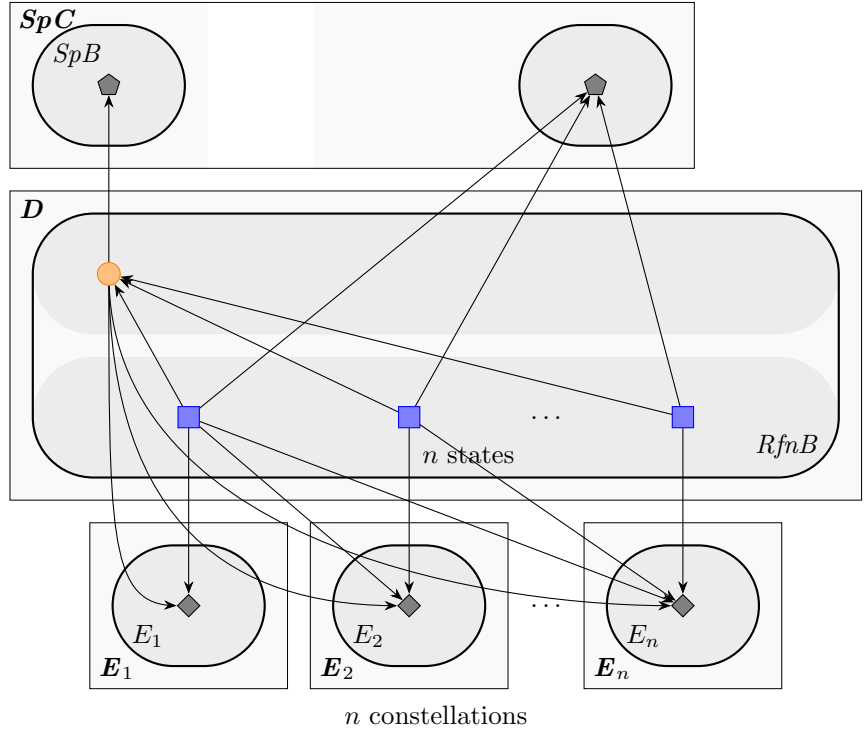
\begin{figure}[tb]
\caption{$\textsc{PostprocessNewBottom}$ is too slow.\label{fig:postprocess:slow}}
\begin{center}
\begin{tikzpicture}[>=Stealth]
	\fill[gray!5]
		(1.9,6.9) rectangle (4.5,9.1)
		(5.9,6.9) rectangle (10.9,9.1);
	\draw[very thin]
		(1.9,9.1) node[below right,inner sep=1mm]{$\culpritcst$} rectangle (10.9,6.9);
	\draw[very thin,fill=gray!5]
		(2.95,0)node[above right]{$\cst{E}_1$}	rectangle (5.55,2.2)
		(5.85,0)node[above right]{$\cst{E}_2$}	rectangle (8.45,2.2)
		(9.45,0)node[above right]{$\cst{E}_n$}	rectangle (12.05,2.2)
		(1.9,6.6)node[below right]{$\cst{D}$}	rectangle (13.1,2.5);
	\draw	(7.5,-0.1)	node[below]{$n$ constellations};

	\fill[gray!15,rounded corners=8mm]
		(2.2,4.7) rectangle (12.8,6.3)
		(2.2,2.8) rectangle (12.8,4.4);
	\draw[thick,rounded corners=8mm]
		(2.2,6.3) rectangle (12.8,2.8)	node[above left,inner sep=1.7mm+2pt]{$\splittedblk$};
	\draw[thick,fill=gray!15,rounded corners=8mm]
		(2.2,8.8) node[below right,inner sep=1.7mm+2pt]{$\culpritblk$}	rectangle (4.2,7.2)
		(8.6,7.2) rectangle (10.6,8.8)

		(3.25,0.3) node[above right,inner sep=1.7mm+2pt]{$E_1$}	rectangle (5.25,1.9)
		(6.15,0.3) node[above right,inner sep=1.7mm+2pt]{$E_2$}	rectangle (8.15,1.9)
		(9.75,0.3) node[above right,inner sep=1.7mm+2pt]{$E_n$}	rectangle (11.75,1.9);

	\node[minimum size=3mm,inner sep=0pt,diamond,draw,fill=black!50] at (4.25,1.1) (bot1){};
	\node[minimum size=3mm,inner sep=0pt,diamond,draw,fill=black!50] at (7.15,1.1) (bot2){};
	\node[inner sep=1pt] at (8.95,1.1) {$\ldots$};
	\node[minimum size=3mm,inner sep=0pt,diamond,draw,fill=black!50] at (10.75,1.1) (botn){};
	\draw[->]	(4.25,3.6) to (bot1);
	\draw[->]	(4.25,3.6) to (bot2);
	\draw[->]	(4.25,3.6) to (botn);
	\draw[->]	(7.15,3.6) to (bot2);
	\draw[->]	(7.15,3.6) to (botn);
	\draw[->]	(10.75,3.6) to (botn);
	\draw[->]	(3.2,5.5) .. controls (3.2,2) and (3.2,1.1) .. (bot1);
	\draw[->]	(3.2,5.5) .. controls (3.2,2) and (5,1.1) .. (bot2);
	\draw[->]	(3.2,5.5) .. controls (3.2,2) and (8,1.1) .. (botn);

	\node[minimum size=3mm,inner sep=0pt,regular polygon,draw=black,fill=black!50] at (3.2,8) (orangegoal){};
	\draw[->,rounded corners=5mm]	(3.2,5.5) to (orangegoal);
	\node[minimum size=3mm,inner sep=0pt,regular polygon,draw=black,fill=black!50] at (9.6,8) (bluegoal){};
	\draw[->]	(4.25,3.6) to (bluegoal);
	\draw[->]	(7.15,3.6) to (bluegoal);
	\draw[->]	(10.75,3.6) to (bluegoal);

	\node[minimum size=3mm,inner sep=0pt,circle,draw=orange,fill=orange!50] at (3.2,5.5) (r){};
	\draw[->]	(4.25,3.6) to (r);
	\draw[->]	(7.15,3.6) to (r);
	\draw[->]	(10.75,3.6) to (r);

	\node[minimum size=2.8mm,inner sep=0pt,rectangle,draw=blue,fill=blue!50] at (4.25,3.6) (blue1){};
	\node[minimum size=2.8mm,inner sep=0pt,rectangle,draw=blue,fill=blue!50] at (7.15,3.6) (blue2){};
	\node[inner sep=1pt] at	(8.95,3.6) (bluedots){$\ldots$};
	\node[minimum size=2.8mm,inner sep=0pt,rectangle,draw=blue,fill=blue!50] at (10.75,3.6) (bluen){};
	\node[right] at (7.2,3.1){$n$ states};
\end{tikzpicture}
\end{center}
\end{figure}

Assume that the partition in Figure~\ref{fig:postprocess:slow} has been reached.
Then, we choose $\culpritblk$ as splitter.
the \orangecirc-state is a (weak) predecessor of $\culpritblk$, but not of $\culpritcst \setminus \culpritblk$,
and therefore is split off from the remainder of $\splittedblk$.
This turns all \bluesquare-states into new bottom states.
First, it is registered that $(\splittedblk, \cst{E}_1)$, \ldots, $(\splittedblk, \cst{E}_n)$ all need postprocessing.
(Also $(\splittedblk, \culpritcst)$ needs postprocessing, but we will disregard it in the lower bound for timing.)
Then, one for one, these pairs are handled.
Suppose it starts with $(\splittedblk, \cst{E}_1)$.
The algorithm will find that it has to split $\splittedblk$ into two parts,
namely the \bluesquare-state that is a predecessor of $\cst{E}_1$ and the $n-1$ other \bluesquare-states.
Then, it walks over the $n-1$ pairs $(\splittedblk, \cst{E}_i)$ that still need postprocessing
and their lists $S_{\cst{E}_i}$, containing altogether $2 + 3 + \cdots + n$ states;
from each list, it will remove the first \bluesquare-state.
In total, $\frac{1}{2} n (n-1) - 1$ list entries are read or removed.
After that, the algorithm may handle $(\splittedblk, \cst{E}_2)$,
split off one more \bluesquare-state from the rest,
and walk over the $n-2$ remaining pairs $(\splittedblk, \cst{E}_i)$.
Here, $\frac{1}{2} (n-1)(n-2) - 1$ list entries are read or removed.
For all the pairs up to $(\splittedblk, \cst{E}_n)$, the algorithm reads and finally removes $\Theta\left(n^3\right)$ list entries.

The Kripke structure in Figure~\ref{fig:postprocess:slow} has $\bigo{n}$ states and $\bigo{n^2}$ transitions.
Therefore, the algorithm should run in time $\bigo{n^2 \log n}$.
However, it takes $\bigOmega{n^3}$ time.
When we think of variants of this Kripke structure
(e.\,g.\@ reduced outdegree of the \orangecirc-state,
or multiple \bluesquare-states with a transition to the same \blackdiamond-state),
we find that there are actually $d$ iterations over $\Theta \left(m\right)$ states.

\begin{lemma}
	\label{lem:timebound2}
	Postprocessing new bottom states as described in \cite{GrooteW16}
	has a worst-case time complexity of $\bigOmega{m d}$.
\end{lemma}

\subsection{A faster $\text{\sc PostprocessNewBottom}$}

The main idea for correcting the time bound was already hinted at earlier:
Lines~\ref{alg:newattention-begin}--\ref{alg:newattention-end}
try to distribute $S_{\cst{C}}$ over $\newblk$ and what remains of $\hat{B}$
in time proportional to the outgoing transitions of $\newblk$,
while keeping the order of $S_{\cst{C}}$ in line with the order of $\newbottomstates$.
This can be achieved
if one distributes $S_{\cst{C}}$ simultaneously
with distributing the states in $\hat{B}$ and their outgoing transitions themselves
in Line~\ref{alg:trysplitprime:movetonewblk}.
If all else fails, even constructing $S_{\cst{C}}$ for $\newblk$ from scratch
can fit the time bound $\bigo{\lvert \mathit{out}(\newblk) \rvert}$.

\newlength{\lw}\setlength{\lw}{0.8pt}
\begin{figure}[tbp]
\caption{$\textsc{PostprocessNewBottom}$ is still too slow.\label{fig:postprocess:stilltooslow}}
\begin{center}
\begin{tikzpicture}[>=Stealth]
	\draw[every node/.style={rounded rectangle,draw,fill=white,inner sep=4pt,anchor=mid}]
		(-2,  1   ) node (s1){\makebox[9pt]{$s_1$}}
		(-2,- 1   ) node (s2){\makebox[9pt]{$s_2$}}
		(-2,- 3.75) node (s3){\makebox[9pt]{$s_3$}}
		(-2,- 7.25) node (s4){\makebox[9pt]{$s_4$}}
		( 0,  0   ) node(s10){\makebox[9pt]{$s_{10}$}}
		( 0,- 2   ) node(s20){\makebox[9pt]{$s_{20}$}}
		( 0,- 2.75) node(s21){\makebox[9pt]{$s_{21}$}}
		( 0,- 4.75) node(s30){\makebox[9pt]{$s_{30}$}}
		( 0,- 5.5 ) node(s31){\makebox[9pt]{$s_{31}$}}
		( 0,- 6.25) node(s32){\makebox[9pt]{$s_{32}$}}
		( 0,- 8.25) node(s40){\makebox[9pt]{$s_{40}$}}
		( 0,- 9   ) node(s41){\makebox[9pt]{$s_{41}$}}
		( 0,- 9.75) node(s42){\makebox[9pt]{$s_{42}$}}
		( 0,-10.5 ) node(s43){\makebox[9pt]{$s_{43}$}}
		(-1,-11.5) node(t4)[draw=black!20]  {\makebox[9pt]{\vphantom{$s_0$}}}
		++(-2\lw,-\lw) node(t3)[draw=black!40]  {\makebox[9pt]{\vphantom{$s_0$}}}
		++(-2\lw,-\lw) node(t2)[draw=black!60]  {\makebox[9pt]{\vphantom{$s_0$}}}
		++(-2\lw,-\lw) node(t1)[draw=black!80]  {\makebox[9pt]{\vphantom{$s_0$}}}
		(-4,-13) node (sb){\makebox[9pt]{$s_\mathrm{b}$}}
	;
	\draw[rounded corners=3mm,line width=4.92\lw] (t2.west) -| ($(sb.north)+( 1.96\lw,12\lw)$);
	\draw[every node/.style={rounded rectangle,draw,fill=white,inner sep=4pt,anchor=mid}]
		(-1,-11.5) ++(-8\lw,-4\lw) node(t0)  {\makebox[9pt]{\smash{$t_i$}\vphantom{$s_0$}}}
	;
	\draw[every node/.style={rectangle,draw,very thin,minimum height=8mm,minimum width=1cm}]
		( 1.5,-14.5) node(C4){$\cst{C}_4$}
		( 3  ,-14.5) node(C3){$\cst{C}_3$}
		( 4.5,-14.5) node(C2){$\cst{C}_2$}
		( 6  ,-14.5) node(C1){$\cst{C}_1$}
		( 7.5,-14.5) node(C0){$\cst{C}_0$}
	;
	\begin{scope}[<-]
		\draw[line width=   \lw] (s1) -- ++(0,-0.5);
		\draw[line width=1.5\lw] (s2) -- ++(0,-0.5);
		\draw[line width=2  \lw] (s3) -- ++(0,-0.5);
		\draw[line width=2.5\lw] (s4) -- ++(0,-0.5);
		\draw[line width=5  \lw] (sb) -- ++(0, 0.5);
		\draw[line width=3.5\lw] (C4) -- ++(0,1);
		\draw[line width=3.5\lw] (C3) -- ++(0,1);
		\draw[line width=4  \lw] (C2) -- ++(0,1);
		\draw[line width=5  \lw] (C1) -- ++(0,1);
		\draw[line width=6  \lw] (C0) -- ++(0,1);
	\end{scope}
	\begin{scope}[rounded corners=3mm,line width=\lw,every node/.style={rectangle,fill=white,inner sep=0pt,rounded corners=0pt,line width=0pt,minimum width=1cm,above right}]
		\draw ($ (s1.east)+(-0.2pt, 0      )$) -|                                               ($(C0.north)+( 4.9 \lw,14\lw)$);
		\draw ($(s10.east)+(-0.2pt, 0      )$) -|                                               ($(C1.north)+( 3.92\lw,12\lw)$);
		\draw ($ (s2.east)+(-0.2pt,-0.49\lw)$) -| node[minimum height=2\lw                  ]{} ($(C1.north)+( 2.94\lw,10\lw)$);
		\draw ($ (s2.east)+(-0.2pt, 0.49\lw)$) -|                                               ($(C0.north)+( 3.92\lw,12\lw)$);
		\draw ($(s20.east)+(-0.2pt,-0.49\lw)$) -| node[minimum height=2\lw,minimum width=2cm]{} ($(C2.north)+( 2.94\lw,10\lw)$);
		\draw ($(s20.east)+(-0.2pt, 0.49\lw)$) -|                                               ($(C0.north)+( 2.94\lw,10\lw)$);
		\draw ($(s21.east)+(-0.2pt, 0      )$) -|                                               ($(C2.north)+( 1.96\lw, 7\lw)$);
		\draw ($ (s3.east)+(-0.4pt,-0.98\lw)$) -| node[minimum height=3\lw                  ]{} ($(C2.north)+( 0.98\lw, 5\lw)$);
		\draw ($ (s3.east)+(-0.2pt, 0      )$) -| node[minimum height=2\lw                  ]{} ($(C1.north)+( 1.96\lw, 7\lw)$);
		\draw ($ (s3.east)+(-0.4pt, 0.98\lw)$) -|                                               ($(C0.north)+( 1.96\lw, 7\lw)$);
		\draw ($(s30.east)+(-0.4pt,-0.98\lw)$) -| node[minimum height=3\lw,minimum width=2cm]{} ($(C3.north)+( 2.45\lw, 8\lw)$);
		\draw ($(s30.east)+(-0.2pt, 0      )$) -| node[minimum height=2\lw                  ]{} ($(C1.north)+( 0.98\lw, 5\lw)$);
		\draw ($(s30.east)+(-0.4pt, 0.98\lw)$) -|                                               ($(C0.north)+( 0.98\lw, 5\lw)$);
		\draw ($(s31.east)+(-0.2pt,-0.49\lw)$) -| node[minimum height=2\lw,minimum width=4cm]{} ($(C3.north)+( 1.47\lw, 6\lw)$);
		\draw ($(s31.east)+(-0.2pt, 0.49\lw)$) -|                                               ($(C0.north)+( 0      , 2\lw)$);
		\draw ($(s32.east)+(-0.2pt, 0      )$) -|                                               ($(C3.north)+( 0.49\lw, 3\lw)$);
		\draw ($ (s4.east)+(-0.4pt,-1.47\lw)$) -| node[minimum height=4\lw                  ]{} ($(C3.north)+(-0.49\lw, 3\lw)$);
		\draw ($ (s4.east)+(-0.2pt,-0.49\lw)$) -| node[minimum height=3\lw                  ]{} ($(C2.north)+( 0      , 2\lw)$);
		\draw ($ (s4.east)+(-0.2pt, 0.49\lw)$) -| node[minimum height=2\lw                  ]{} ($(C1.north)+( 0      , 2\lw)$);
		\draw ($ (s4.east)+(-0.4pt, 1.47\lw)$) -|                                               ($(C0.north)+(-0.98\lw, 5\lw)$);
		\draw ($(s40.east)+(-0.4pt,-1.47\lw)$) -| node[minimum height=4\lw,minimum width=2cm]{} ($(C4.north)+( 2.45\lw, 8\lw)$);
		\draw ($(s40.east)+(-0.2pt,-0.49\lw)$) -| node[minimum height=3\lw                  ]{} ($(C2.north)+(-0.98\lw, 5\lw)$);
		\draw ($(s40.east)+(-0.2pt, 0.49\lw)$) -| node[minimum height=2\lw                  ]{} ($(C1.north)+(-0.98\lw, 5\lw)$);
		\draw ($(s40.east)+(-0.4pt, 1.47\lw)$) -|                                               ($(C0.north)+(-1.96\lw, 7\lw)$);
		\draw ($(s41.east)+(-0.4pt,-0.98\lw)$) -| node[minimum height=3\lw,minimum width=4cm]{} ($(C4.north)+( 1.47\lw, 6\lw)$);
		\draw ($(s41.east)+(-0.2pt, 0      )$) -| node[minimum height=2\lw                  ]{} ($(C1.north)+(-1.96\lw, 7\lw)$);
		\draw ($(s41.east)+(-0.4pt, 0.98\lw)$) -|                                               ($(C0.north)+(-2.94\lw,10\lw)$);
		\draw ($(s42.east)+(-0.2pt,-0.49\lw)$) -| node[minimum height=2\lw,minimum width=5cm]{} ($(C4.north)+( 0.49\lw, 3\lw)$);
		\draw ($(s42.east)+(-0.2pt, 0.49\lw)$) -|                                               ($(C0.north)+(-3.92\lw,12\lw)$);
		\draw ($(s43.east)+(-0.2pt, 0      )$) -|                                               ($(C4.north)+(-0.49\lw, 3\lw)$);
		\draw ($ (t0.east)                  $) -| node[minimum height=5\lw                  ]{} ($(C4.north)+(-1.47\lw, 6\lw)$);
		\draw ($ (t1.east)                  $) -| node[minimum height=4\lw                  ]{} ($(C3.north)+(-1.47\lw, 6\lw)$);
		\draw ($ (t2.east)                  $) -| node[minimum height=3\lw                  ]{} ($(C2.north)+(-1.96\lw, 7\lw)$);
		\draw ($ (t3.east)                  $) -| node[minimum height=2\lw                  ]{} ($(C1.north)+(-2.94\lw,10\lw)$);
		\draw ($ (t4.east)                  $) -|                                               ($(C0.north)+(-4.9 \lw,14\lw)$);
		\draw ($ (sb.east)+(-0.4pt,-1.96\lw)$) -| node[minimum height=5\lw                  ]{} ($(C4.north)+(-2.45\lw, 8\lw)$);
		\draw ($ (sb.east)+(-0.4pt,-0.98\lw)$) -| node[minimum height=4\lw                  ]{} ($(C3.north)+(-2.45\lw, 8\lw)$);
		\draw ($ (sb.east)+(-0.2pt, 0      )$) -| node[minimum height=3\lw                  ]{} ($(C2.north)+(-2.94\lw,10\lw)$);
		\draw ($ (sb.east)+(-0.4pt, 0.98\lw)$) -| node[minimum height=2\lw                  ]{} ($(C1.north)+(-3.92\lw,12\lw)$);
		\draw[->] ($(sb.east)+(-0.4pt, 1.96\lw)$) -| (C0.north -| C0.west);

		\draw ($(s1.west)+( 0.2pt, 0      )$) -| ($(sb.north)+(-3.92\lw,12\lw)$);
		\draw ($(s2.west)+( 0.2pt, 0      )$) -| ($(sb.north)+(-2.94\lw,10\lw)$);
		\draw ($(s3.west)+( 0.2pt, 0      )$) -| ($(sb.north)+(-1.96\lw, 7\lw)$);
		\draw ($(s4.west)+( 0.2pt, 0      )$) -| ($(sb.north)+(-0.98\lw, 5\lw)$);

		\draw ($(s10.west)+(0.2pt, 0      )$) -| ($(s1.south)+( 0      ,-2\lw)$);
		\draw ($(s20.west)+(0.2pt, 0      )$) -| ($(s2.south)+( 0.49\lw,-3\lw)$);
		\draw ($(s21.west)+(0.2pt, 0      )$) -| ($(s2.south)+(-0.49\lw,-3\lw)$);
		\draw ($(s30.west)+(0.2pt, 0      )$) -| ($(s3.south)+( 0.98\lw,-5\lw)$);
		\draw ($(s31.west)+(0.2pt, 0      )$) -| ($(s3.south)+( 0      ,-2\lw)$);
		\draw ($(s32.west)+(0.2pt, 0      )$) -| ($(s3.south)+(-0.98\lw,-5\lw)$);
		\draw ($(s40.west)+(0.2pt, 0      )$) -| ($(s4.south)+( 1.47\lw,-6\lw)$);
		\draw ($(s41.west)+(0.2pt, 0      )$) -| ($(s4.south)+( 0.49\lw,-3\lw)$);
		\draw ($(s42.west)+(0.2pt, 0      )$) -| ($(s4.south)+(-0.49\lw,-3\lw)$);
		\draw ($(s43.west)+(0.2pt, 0      )$) -| ($(s4.south)+(-1.47\lw,-6\lw)$);
	\end{scope}
\end{tikzpicture}
\end{center}
\end{figure}
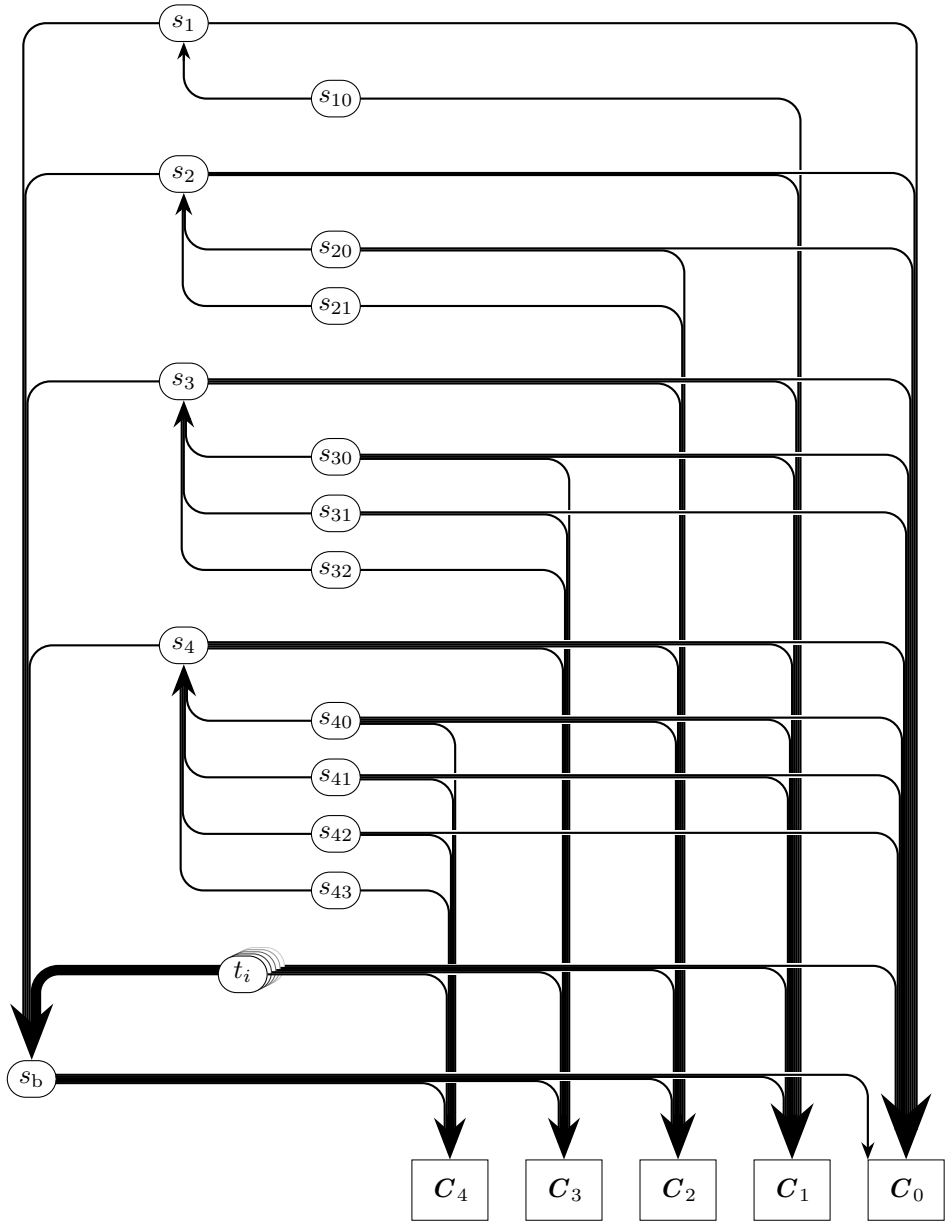

\subsection{$\text{\sc PostprocessNewBottom}$ is still too slow}

The algorithm $\textsc{TrySplit}'$ walks through \emph{all} new bottom states to look for blue states:
in Line~\ref{alg:trysplit:newbottom},
we call it with $\mathit{MaybeBlue} = \newbottomstates$.
There may be a problem in a Kripke structure with a large number of new bottom states
when we have to split over and over again for the same constellation,
because every now and then, another new bottom state is discovered.

Consider the Kripke structure drawn partially in Figure~\ref{fig:postprocess:stilltooslow}.
The states drawn individually are all in the same block,
which is considered for refinement with $\culpritblk \subset \cst{C}_0$ as splitter.
We assume that state $s_\mathrm{b}$ has a transition to $\culpritblk$,
and all other transitions to $\cst{C}_0$ go to $\cst{C}_0 \setminus \culpritblk$.
Before splitting, $s_\mathrm{b}$ is the only bottom state in the block;
however, as soon as it is split off, $s_1, s_2, \ldots$ become new bottom states.

If we extend this construction until we reach $s_n$ and $s_{n,n-1}$,
it will have $\bigo{n^2}$ states
and $\bigo{n^3}$ transitions.
For a better overview, we present the refinements in two levels;
the algorithm follows the order of presentation, but does so iteratively.
\begin{itemize}\setlength{\itemsep}{0pt}
\item	Let's start with $\culpritcst = \cst{C}_0$,
	after splitting it into $\culpritblk$ and the rest:
	this refinement splits $s_\mathrm{b}$ from the rest of the block
	and registers that $s_1, s_2, \ldots$ are new bottom states.
	As a result, the algorithm calls $\textsc{PostprocessNewBottom}$.
\item	Let's refine with $\cst{C}_1$:
	this will split $s_1$ from the block and create another new bottom state $s_{10}$.
	So we have to check for $\cst{C}_0$ again;
	this will now split off $s_{10}$ from the rest.
\item	Then refine with $\cst{C}_2$:
	this splits $s_2$ from the block and creates two more new bottom states $s_{20}$ and $s_{21}$.
	We now have to refine for $\cst{C}_0$ and $\cst{C}_1$ again:
	first $s_{21}$ is split off and then $s_{20}$.
\item	Then we refine with $\cst{C}_3$:
	this will split $s_3$ from the block and create three more new bottom states $s_{30}, s_{31}$ and $s_{32}$.
	Now we can refine for $\cst{C}_0, \cst{C}_1$ and $\cst{C}_2$ again;
	every time, we will split one of these new bottom states off.
\item	etc.
\end{itemize}
The reader will see that in all these refinements, the set of blue states is very small;
however, it is never empty.
We are allowed to spend time proportional to the number of these blue states and their transitions.
In total, there will be about $\Theta\left(n^2\right)$ such refinements.
This on itself is not yet problematic.
However, we can without increasing the complexity class of the Kripke structure,
add another $\bigo{n^2}$ new bottom states $t_1, \ldots, t_{n^2}$ to this block with transitions to each of the $\cst{C}_i$.
Every time we look for the blue states,
we have to run through all new bottom states
to find the states that are not in $S_{\cst{C}_i}$.
So, in reality, each refinement will spend time $\Omega\left(n^2\right)$.
However, as there were only $\bigo{n^3}$ transitions, this is too much.

\subsection{Possible correction}

We should make sure that first those new bottom states are handled that have been found last:
When we call $\textsc{TrySplit}'$ in Line~\ref{alg:trysplit:newbottom},
the blue coroutine first walks through the newest bottom states.
Then, as soon as we know that all blue new bottom states have been found
(based on the difference between the length of $S_{\cst{C}_i}$ and the total number of new bottom states),
we can stop running through the other new bottom states.
As a consequence, the new bottom states found earlier (and possibly already run through earlier) will not be visited another time;
the time spent to skip over states in $\mathit{MaybeBlue} \setminus \mathit{Blue}$ (in Line~\ref{alg:trysplitprime:bluetest}$\ell$),
over all refinements together,
will be bounded by the number of outgoing transitions of new bottom states.

Another solution might be to handle all refinements w.\,r.\,t. a fixed set of new bottom states before considering more new bottom states.
In terms of the original data structure, that would mean to keep the sets $X_B, X_{B'}$ untouched until all refinements are done,
and only after that handle the new bottom states contained therein.

\section{Conclusion}

We showed that the algorithm of \cite{GrooteW16} for computing stuttering equivalence does not meet the acclaimed bound of $\bigo{m \log n}$; instead, there are examples showing it required $\bigOmega{m d}$ time with $d$ the maximum outdegree in the Kripke structure.

Presentation of the algorithm in pseudocode enabled us to identify the parts of the algorithm that are responsible for the overrun of the time bound. We are convinced that, after correcting $\textsc{TrySplit}'$ and $\textsc{PostprocessNewBottom}$,
the time budgets (as indicated in the pseudocode) are met.
Therefore, we are emboldened to confirm the main result of \cite{GrooteW16}:

\begin{theorem}\label{thm:correctupperbound}
	It is possible to calculate the stuttering equivalence of a Kripke structure
	in $\bigo{m \log n}$ time
	and $\bigo{m}$ memory
	(by using the corrected Algorithm~\ref{algo:feb11}).
\end{theorem}

\bibliographystyle{plain}
\bibliography{algorithm}

\end{document}